\documentclass[letterpaper,conference]{ieeeconf}
\usepackage{booktabs}
\usepackage{graphicx}
\usepackage{amsmath,amsfonts,graphicx}
\usepackage{svg}
\usepackage{booktabs}
\usepackage{comment}
\usepackage{nidanfloat}
\usepackage{algorithmic}
\usepackage{array}
\usepackage{url}
\usepackage{balance}
\usepackage{hyperref}
\usepackage{array}
\usepackage{multirow}
\usepackage{xspace}
\usepackage{cite}

\IEEEoverridecommandlockouts
\begin{document}

\title{\LARGE \bf Disentangled Adversarial Transfer Learning for Physiological Biosignals}

\author{Mo Han$^{1}$, Ozan {\"O}zdenizci$^{1}$, Ye Wang$^{2}$, Toshiaki Koike-Akino$^{2}$ and Deniz Erdo{\u{g}}mu{\c{s}}$^{1}$%
\thanks{$^{1}$Cognitive Systems Laboratory, Department of Electrical and Computer Engineering, Northeastern University, Boston, MA 02115, USA, E-mail: \{han, oozdenizci, erdogmus\}@ece.neu.edu.}
\thanks{$^{2}$Mitsubishi Electric Research Laboratories (MERL), Cambridge, MA 02139, USA, E-mail: \{yewang, koike\}@merl.com.}%
\thanks{M.~Han was an intern at MERL during this work. O.~{\"O}zdenizci and D.~Erdo{\u{g}}mu{\c{s}} are partially supported by NSF (IIS-1149570, CNS-1544895, IIS-1715858), DHHS (90RE5017-02-01), and NIH (R01DC009834).}%
}

\maketitle

\begin{abstract}
Recent developments in wearable sensors demonstrate promising results for monitoring physiological status in effective and comfortable ways. One major challenge of physiological status assessment is the problem of transfer learning caused by the domain inconsistency of biosignals across users or different recording sessions from the same user. We propose an adversarial inference approach for transfer learning to extract disentangled nuisance-robust representations from physiological biosignal data in stress status level assessment. We exploit the trade-off between task-related features and person-discriminative information by using both an adversary network and a nuisance network to jointly manipulate and disentangle the learned latent representations by the encoder, which are then input to a discriminative classifier. Results on cross-subjects transfer evaluations demonstrate the benefits of the proposed adversarial framework, and thus show its capabilities to adapt to a broader range of subjects. Finally we highlight that our proposed adversarial transfer learning approach is also applicable to other deep feature learning frameworks.
\end{abstract}

\begin{keywords}
stress level assessment, physiological biosignals, adversarial networks, transfer learning, deep neural networks, disentangled representation learning
\end{keywords}

\section{Introduction}

Novel designs of wearable sensors demonstrate promising results for monitoring physiological status (e.g., stress level assessment) in humans. A traditional method to assess such activity was by measuring electroencephalography (EEG) \cite{petrantonakis2009emotion}. However, EEG-based monitoring requires either surface (non-invasive) or implanted (invasive) electrodes and frequent calibration to account for sensor sensitivity to external factors, which increases system expense and decreases user comfort. Usage of non-EEG physiological biosignals \cite{NonEEG1,NonEEG2,NonEEG3,giakoumis2013subject,giannakakis2019review,ozdenizci2018time} avoids the aforementioned issues with a wrist-worn platform in more effective, comfortable, and less expensive ways.

One major issue in identifying different physiological states is to prevent undesired variability among different subjects or different recording sessions from a single subject. Generally, given the fact that most biosignal datasets are of smaller scale, transfer learning \cite{trans2,trans3,trans4} aims to cope with the change in data distributions, in order to process data from a wider range of users. Notably, promising results were demonstrated in transfer learning by censoring learned discriminative representations within an adversarial training scheme \cite{advTrans1,advTrans2,advTrans3,ozdenizci2020learning,advTrans4,advTrans-csl2}. Adversarial representation learning can allow the representation to predict dependent variables while simultaneously taking advantage of an adaptive measure to control the extent of its dependency during training.

In this study, we propose an adversarial inference approach for transfer learning to exploit disentangled nuisance-robust representations from physiological biosignal data in stress status level decoding. Particularly different from common deep learning frameworks, we exploit a trade-off between task-related features and person-discriminative information by using additional censoring network blocks to manipulate the learned latent representations using adversarial training schemes. By jointly training the adversary, nuisance and classifier units, task-discriminative features are incorporated into the final prediction, while simultaneously the biosignal characteristics from new users could also be projected to local features extracted from the existing subject pool for reference purposes, especially when new users demonstrate similar biosignal behaviors to the training set subjects. We perform empirical assessments on a publicly available dataset with extensive parameter explorations. Results demonstrate the advantage of our disentangled adversarial transfer learning framework with a proof of concept through cross-subject evaluations. Moreover, we highlight that the proposed adversarial transfer learning framework is applicable to other deep learning network approaches that are available, depending on the characteristics of the signal to be learned.

\section{Methods}

\subsection{Disentangled Adversarial Transfer Learning}

Let $\{(X_i,y_i,s_i)\} _{i=1}^{n}$ denote the training data set, where $X_i \in \mathbb{R}^{C \times T}$ is the raw data matrix at trial $i$ recorded from $C$ dimensions for $T$ discretized time samples, $y_i \in {\{0,1,...,L-1\}}$ is the label of corresponding user stress level status or task among $L$ categories, and $s_i \in {\{1,...,S-1,S\}} $ indicates the subject identification (ID) number from whom the data was recorded among $S$ individuals. Here we assume the task/status $y$ and subject ID $s$ are marginally independent, and the data is generated dependently on $y$ and $s$ jointly i.e., $X\sim p(X|y,s)$. Our aim is to build a discriminative model which can classify/predict the category $y$ given observation $X$, where the model is generalized across subjects and invariant to the variability in subject IDs $s$, which is regarded as a nuisance variable involved in the data generation process.

\begin{figure*}[t]
  \centering
  \includegraphics[clip, trim=0cm 0cm 9cm 3cm,width=0.79\textwidth]{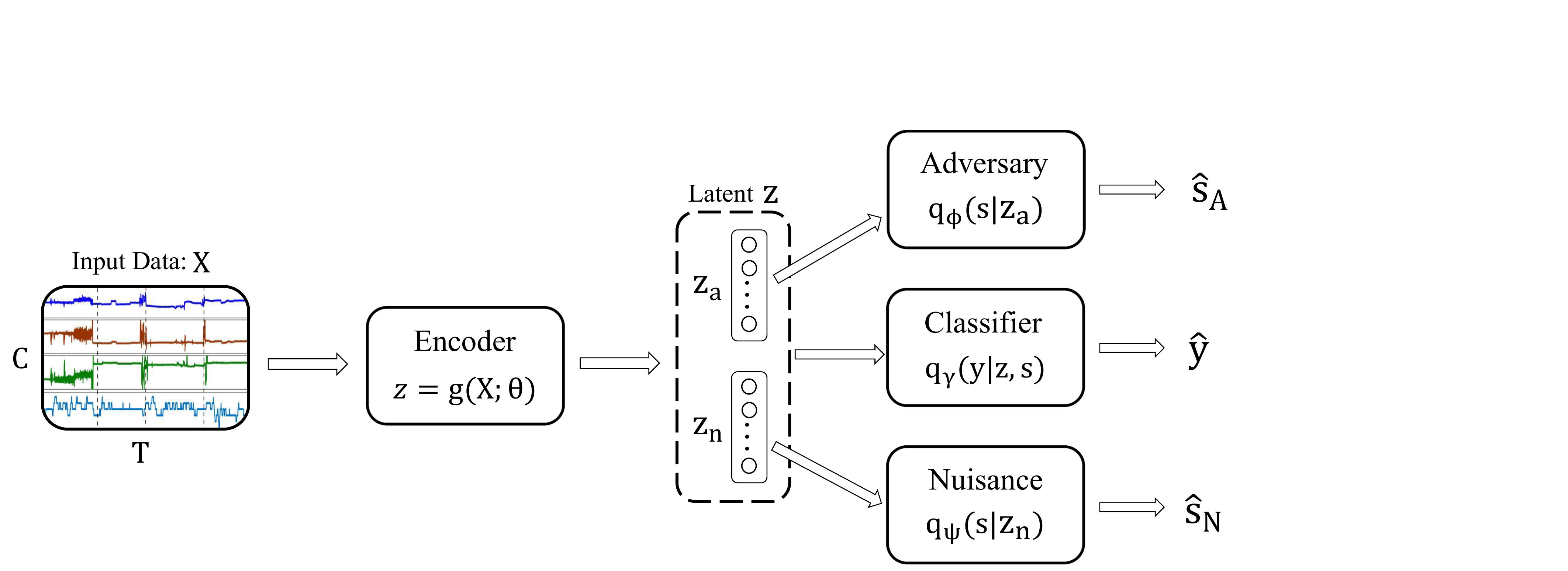}\vspace{0.2cm}
  \caption{A deterministic encoder $z=g(X;\theta)$ with parameters $\theta$ is trained to learn the overall latent representation $z$ from data $X$, where latent $z$ consists of two sub-parts: $z_a$ and $z_n$, based on a ratio of $(1-r_N):r_N$ over its dimensionality. The latent sub-component $z_a$ is used as input to an \textit{adversary network} with parameters $\phi$, while $z_n$ is used as input to a \textit{nuisance network} parameterized by $\psi$. Both $z_a$ and $z_n$ (i.e., $z=[z_a,z_n]$) are together used as input for the \textit{main classifier network} with parameters $\gamma$, alongside the condition $s$.}
\vspace{-0.25cm}
\label{model}
\end{figure*}

In the proposed framework, a deterministic encoder $z=g(X;\theta)$ with parameters $\theta$ is trained to learn the latent representation $z$ from data $X$, where the latent $z$ consists of two sub-parts: $z_a$ and $z_n$, based on a ratio of $(1-r_N):r_N$ over its dimensionality. The latent sub-part $z_a$ is used as the input to an \textit{adversary network} with parameters $\phi$, while $z_n$ serves as the input to a \textit{nuisance network} parameterized by $\psi$, as illustrated in Figure~\ref{model}. Complete latent representation $z$ (i.e., concatenation of $z_a$ and $z_n$) is further used as an input to the \textit{main classifier network} with parameters $\gamma$.

In order to filter factors of variation caused by $s$ out of $z_a$, the encoder is forced to minimize the likelihood $q_\phi\left(s|z_a\right)$, while at the same time maximizing the likelihood $q_\psi\left(s|z_n\right)$ to retain sufficient subject-discriminative information within $z_n$. The main classifier network is further conditioned on $s$ alongside latent $z$, and trained towards the main classification task to predict the category label $y$ by maximizing the likelihood $q_\gamma\left(y|g(X;\theta),s\right)$. Overall, we propose the following objective to train the encoder-classifier pair as follows:
\begin{equation}
\label{totalloss}
\begin{split}
          \max\limits_{\theta,\gamma,\psi} \min\limits_{\phi} ~~
          & \mathbb{E} [ \log q_\gamma \left(y|g(X;\theta),s \right)
          + \lambda_N \log q_\psi \left(s|z_n \right)   \\
          & - \lambda_A \log q_\phi \left(s|z_a \right) ]
\end{split}
\end{equation}
where $\lambda_A$ and $\lambda_N$ denote the weight parameters for adversary and nuisance networks respectively, controlled to adjust the trade-off between invariance and identification performance. Setting $\lambda_A = \lambda_N = 0$ indicates training a regular discriminative neural network structure without disentangling the transfer learning units. Note that besides the overall objective, both of the adversary and nuisance networks are also trained separately to predict variable $s$ by maximizing the likelihoods $q_\phi\left(s|z_a\right)$ and $q_\psi\left(s|z_n\right)$ respectively. Neural network weights are optimized by every training data batch via stochastic gradient descent; for each training batch, weights for the adversary network, nuisance network and the classifier network are updated alternatingly according to their corresponding softmax cross-entropy loss.

Disentangling of $z$ into sub-parts $z_a$ and $z_n$ is proposed to systematically re-arrange the distribution of task- and subject- related features. While $z_a$ conceals subject information indicated through $s$, $z_n$ is trained to retain subject-related information within the learned sub-component. By dissociating the nuisance variable $s$ from task-related discriminative features in a more clear way, the model is extrapolated into a broader domain of subjects. For the input data of users unknown to the training subject set, task-related features $z_a$ would be incorporated into the final prediction, whereas the biosignal behaviors which are similar to known subjects could also be projected to $z_n$ to serve as a reference.

\subsection{Model Architecture}

Deep neural networks in biomedical signal processing were recently demonstrated as powerful generic feature extractors \cite{ozdenizci2020learning,advTrans-csl2,atzori2016deep,faust2018deep}. In the view of these progress, each block in the proposed model in Figure~\ref{model} is composed of neural networks for further assessments. It is worth noting that our proposed framework is applicable to any other discriminative representation learning network, depending on the characteristics of the signal of interest.

The encoder consists of two linear layers with 100 units per layer, since deeper layers did not to improve the performance significantly but yet increasing the amount of parameters to be estimated and hence causing possible overfitting to the training data. In our preliminary analyses, we also did not observe significant improvements by altering the number of units at each layer. Representation $z$ with dimension $d=100$ is then generated and split into $z_n$ and $z_a$ with dimensions of $d \cdot r_N$ and $d \cdot (1 - r_N)$ respectively. Attached to the encoder, the adversary network, nuisance network and the main classifier are each built as a single hidden layer multilayer perceptron (MLP) with ReLU nonlinearity. Learned representation sub-parts $z_a$ and $z_n$ are respectively used by adversary and nuisance networks with output dimensionality of $S$ for classification of subject IDs. Similarly complete representation $z$ is used as input to the main classifier network with an output dimensionality of $L$ for task label decoding.

\section{Experimental Evaluation and Results}

\subsection{Physiological Biosignal Dataset}

We perform the experimental evaluations on a publicly available physiological biosignal dataset for the assessment of different stress status levels \cite{NonEEG1}. This database consists of physiological biosignals for inferring 4 different stress status ($L=4$) from 20 healthy subjects ($S=20$), including physical stress, cognitive stress, emotional stress and relaxation. The data was collected by non-invasive wrist worn biosensors and contains electrodermal activity (EDA), temperature, acceleration, heart rate, and arterial oxygen level, where acceleration is composed of data from three channels. Thus the dataset consists of signals from 7 channels in total ($C=7$), which we downsampled to 1 Hz to align all data sources. For each of the stress status states, a corresponding task of 5 minutes (i.e., 300 time samples with $T=300$) was assigned to subjects for inducing the stress levels. Each subject performed a total of 7 trials, where 4 out of the 7 trials were for the relaxation status. To account for imbalanced number of trials across classes, we only used the first trial of the relaxation trials and ignored the rest, resulting in one trial for each of the four stress status levels.

\subsection{Experiment Implementation}

For the model described above, according to the dataset, we have $C=7$, $T=300$, $L=4$ and $S=20$. The parameters to be determined for the disentangled adversarial model were regularization weights $\lambda_A$ and $\lambda_N$, and the rate of nuisance representation $r_N$. An intuitive way to optimize is by a parameter sweep. To perform this, we trained our models with various parameter combinations, and favored the decreases in adversary accuracy with increasing nuisance accuracy, while maintaining a relatively stable accuracy for the main classifier on the validation sets. 

\renewcommand{\arraystretch}{1.2}
\begin{table*}[t!]
    \caption{Cross-subject model evaluation accuracies of three representation learning frameworks: (1) non-adversarial, (2) adversarial, (3) disentangled adversarial, with varying parameter choices. Accuracies for the adversary, nuisance and main classifier networks are presented. To illustrate: in the first row, the non-adversarial model ($\lambda_A=0$) has a classification accuracy of 79.88\%, together with 71.13\% adversary network accuracy and 6.17\% nuisance network accuracy.}
    \centering
    \begin{tabular}{c c c c c c c}
    \toprule
    & $\lambda_A$ & $\lambda_N$ & $r_N$ & \textbf{Main Classifier} & \textbf{Adversary Network} & \textbf{Nuisance Network} \\ \toprule
    Non-Adversarial & 0 & 0 & 0 & 79.88\% & 71.13\% & 6.17\% \\  \midrule
    \multirow{2}{*}{Adversarial} & 0.005 & 0 & 0 & 79.97\% & 35.62\% & 6.15\% \\
    & 0.1 & 0 & 0 & 80.34\% & 8.08\% & 6.20\% \\ \midrule
    \multirow{5}{*}{\textbf{Disentangled Adversarial}} & 0.1 & 0.001 & 0.2 & 80.62\% & 7.05\% & 39.03\% \\
    & \textbf{0.1} & \textbf{0.005} & \textbf{0.2} & \textbf{80.66\%} & \textbf{7.90\%} & \textbf{55.54\%} \\
    & 0.1 & 0.05 & 0.2 & 80.04\% & 7.37\% & 78.83\% \\
    & 0.1 & 0.1 & 0.2 & 80.36\% & 8.08\% & 83.72\% \\
    & 0.1 & 0.2 & 0.2 & 80.22\% & 8.05\% & 87.26\% \\
    \bottomrule
    \end{tabular}
    \label{results}
\end{table*}

\begin{figure}[h!]
  \centering
  \includegraphics[width=0.34\textwidth]{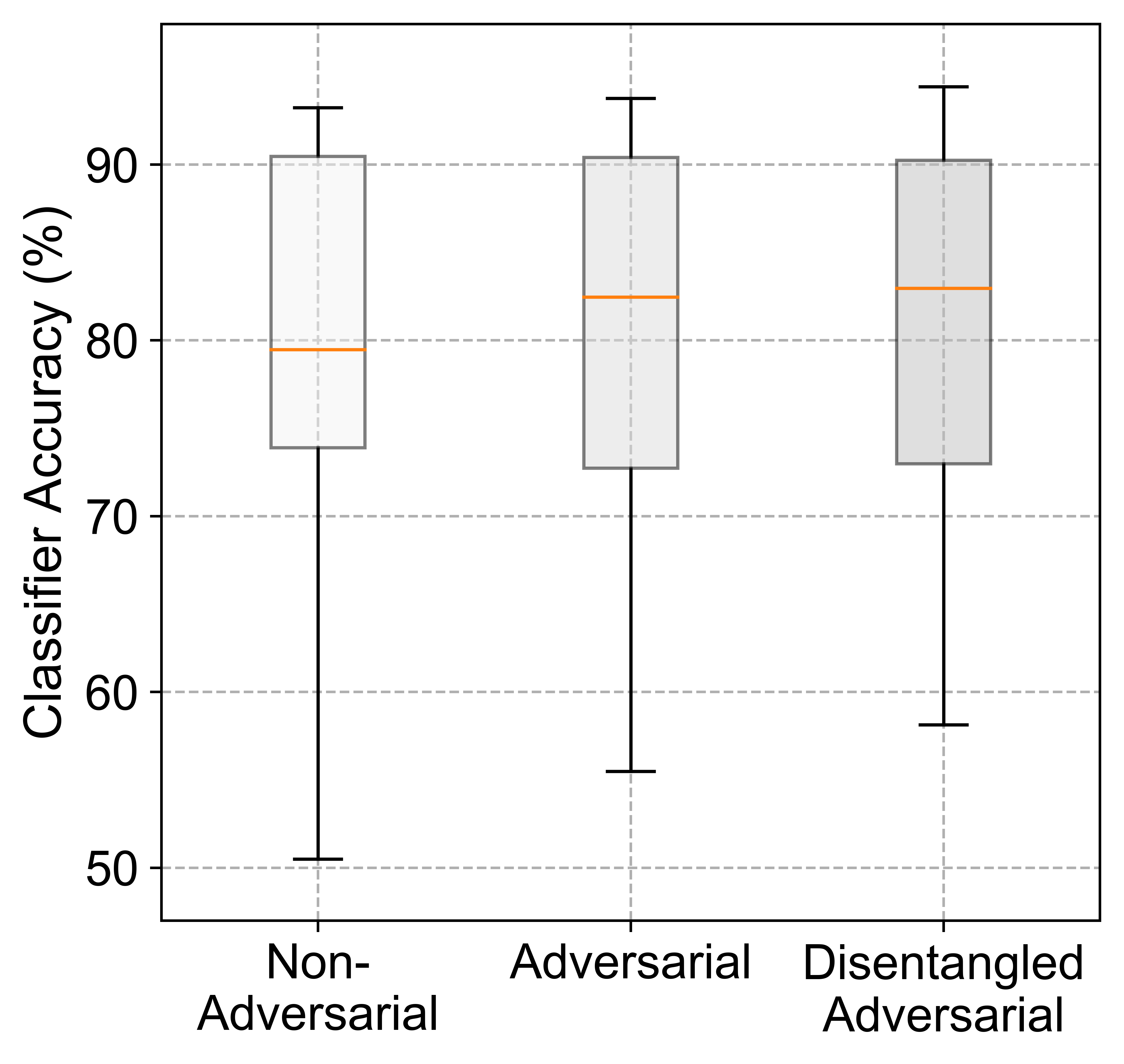}
  \caption{Transfer learning accuracies for 20 held-out subjects on three specific model cases. Central line marks represent the median, upper and lower bounds of the boxes represent first and third quartiles, and dashed lines represent extreme values. For the adversarial model $\lambda_A=0.1$ and $\lambda_N=r_N=0$, whereas for the disentangled adversarial model $\lambda_A=0.1$, $\lambda_N=0.005$ and $r_N=0.2$. With more network units attached to the encoder, we obtain higher accuracies since the proposed transfer learning framework becomes more robust and stable to data from unknown subjects.}
  \vspace{-0.3cm}
\label{results_box}
\end{figure}

To initially reduce the amount of parameter combinations, we first optimized the model for $\lambda_A=0.05$ and $\lambda_A=0.1$ with $\lambda_N=0$ and $r_N=0$, which is the case of \textit{adversarial model} with only adversary network attached. Later, based on the assumption that the subject-related representation $z_n$ accounts for a relatively small proportion among $z$ in order to solve the task-specific problem, we fixed the rate of nuisance representation to $r_N=0.2$. With a fixed $\lambda_A$ and $r_N$, we further assessed the model with varying $\lambda_N \in \{ 0.001, 0.005, 0.05, 0.01, 0.2 \}$, which is the \textit{disentangled adversarial} model with both adversary and nuisance networks attached. It is essential to note that these parameters could still be changed and optimally chosen by cross-validating the model learning stage even further, since for different selections of each parameter which were not covered in this implementation there are corresponding variable combinations to be optimized. Still, the adversarial transfer learning framework could be applied for any other specifications. Evaluations were performed by cross-subjects analyses using a leave-one-subject-out approach, where the left-out subject constituted the cross-subject test set, and the training and validation sets were composed of 90\% and 10\% random trial splits from the remaining subjects.

\subsection{Results and Discussion}

We performed cross-subjects analyses to evaluate the trained models, which is an indicator for transfer learning performances. As shown in Table~\ref{results}, we first assessed the \textit{non-adversarial} models with $\lambda_A=0$, $\lambda_N=0$ and $r_N=0$. Later we evaluated the \textit{adversarial} network with $\lambda_A=0.005$ and $\lambda_A=0.1$ respectively with $\lambda_N=0$ and $r_N=0$ to approximately reduce the number of parameters. Finally, we fixed $\lambda_A=0.1$ and $r_N=0.2$ in order to observe the representation learning capability of the complete \textit{disentangled adversarial} transfer learning model with different choices of $\lambda_N$. For each model we evaluated the accuracy of the main classifier (4-class decoding), as well as the adversary and nuisance networks (20-class decoding). A higher accuracy of main classifier indicates better discrimination of stress status levels, a lower accuracy of adversary network demonstrates that more task-specific information are preserved in the learned representation $z_a$, and a higher accuracy of nuisance network shows that more subject-dependent features are existing in the representation $z_n$. Thus our aim is to keep the accuracy of main classifier stable while decreasing the adversary accuracy and increasing the nuisance accuracy.

In Table~\ref{results} we observe that the non-adversarial model can indeed learn features which yield a status-classification accuracy of 79.88\%, yet with a 71.13\% adversary network accuracy and a 6.17\% nuisance network accuracy. We further notice that with increasing $\lambda_A$, the adversary network accuracy descends dramatically towards chance level and thus more task-discriminative features are exploited by $z_a$, while the main classifier accuracy slightly increases. Specifically, $\lambda_A=0.1$ is more preferable than $\lambda_A=0.005$ in this case. Moreover, under the particular setting of $\lambda_A=0.1$ and $r_N=0.2$, we observe that higher $\lambda_N$ censors the encoder with significantly increased nuisance network accuracies, and therefore enforces stronger extraction of subject information into $z_n$, with slightly higher but relatively stable main classifier accuracies. Figure~\ref{results_box} demonstrates the transfer learning results for the 20 held-out subjects on three specific model training conditions, where we observe that with our approach using both adversary and nuisance network units attached to the encoder, the classifier improves the worst-case accuracies significantly and shows more stable performances across different left-out subjects, since the proposed transfer learning framework becomes more robust to decode data of unknown subjects from a broader range.

\section{Conclusion}

This study proposes a framework for disentangled adversarial transfer learning to extract nuisance-robust representations from physiological biosignal data in stress status level decoding. Different from common deep learning network architectures, in our proposed model, additional adversary and nuisance networks are attached to the output of the feature learning encoder for manipulating the latent representations. We exploit a novel objective towards which the adversary network, nuisance network and the encoder-classifier pair are jointly trained. We perform cross-subject transfer learning evaluations over a publicly available physiological biosignal dataset for stress status level monitoring. Results demonstrate the benefits of the proposed disentangled adversarial framework in transfer learning with input data from novel users, and thus demonstrate better adaptability to a wider range of subjects. Our proposed adversarial transfer learning model is also applicable to any other deep feature learning approach, where the feature encoders could be manipulated accordingly based on different input signal characteristics.



\end{document}